# INFRARED SPECTRA OF DEHYDROGENATED CARBON MOLECULES

Stanislav Kuzmin and W. W. Duley

Department of Physics and Astronomy, University of Waterloo, 200 University Avenue West, Waterloo, ON N2L 3G1, Canada:  wwduley@uwaterloo.ca

ABSTRACT

The detection of fullerene molecules in a variety of astrophysical environments suggests that smaller dehydrogenated carbon molecules may also be present in these sources. One of these is planar $C_{24}$ which has been shown to be more stable than the cage fullerene with the same number of carbon atoms. To facilitate searches for $C_{24}$ and some simple derivatives we have calculated infrared spectra for these molecules using first principles density functional techniques (DFT). Infrared spectra are also presented for several novel carbon cage molecules formed from dehydrogenated polycyclic aromatic hydrocarbon molecules. Infrared spectra of a number of these molecules are quite distinctive and we discuss the possibility of detecting these species in the presence of $C_{60}$ and other fullerenes.

1. INTRODUCTION

The recent detection of $C_{60}$ and $C_{70}$ fullerenes in planetary and reflection nebulae (Cami et al. 2010, Sellgren et al. 2010, Garcia-Hernandez et al. 2010) confirms a long-standing prediction that carbon molecules of this type are present in the interstellar medium (Kroto & Jura 1992, Goeres & Sedlmayr 1992). $C_{60}$ and $C_{70}$ are noted for their unique geometry and chemical stability, and are two examples of a much wider class of fullerene molecules having spherical, ellipsoidal or tubular configurations (Dresselhaus et al. 1996). Smaller fullerenes include $C_{24}$, $C_{26}$, $C_{28}$, etc., but all these molecules are less stable than $C_{60}$ because of strain introduced as a result of the requirement to bend part of a defected graphite sheet to form a 3-dimensional figure. $C_{20}$, the smallest possible fullerene, is the most highly strained of these structures and may be less stable than the bowl-shaped $C_{20}$ isomer (An et al. 2005). High level calculations show that the fullerene form of $C_{24}$ is less stable than that of planar $C_{24}$ which is essentially a dehydrogenated coronene molecule, $C_{24}H_{12}$ (Kent et al. 1998). While buckyball-type fullerenes containing at least one pentagonal ring, are found to be energetically favored for larger molecules, other structures involving arrays of hexagonal rings can be the preferred configuration when the molecule contains fewer than $\approx 30$ carbon atoms. Examples of these are highly stable cage-type carbon molecules based on stacked hexagonal $C_6$ rings (Malcioglu & Erkoc 2005, Kuzmin & Duley 2010). In an astrophysical context, this suggests that a variety of other dehydrogenated carbon clusters could occur along with $C_{60}$ and other fullerenes in astrophysical sources and that these objects may then provide spectral information on new molecular species.

This paper reports some results of a first principles calculation of infrared spectra of a variety of planar and carbon cage molecules based exclusively on conformations of hexagonal rings (dehydrogenated PAHs). These molecules exhibit a number of characteristic spectral features that can be used to guide a search for novel small carbon clusters in astrophysical objects.

## 2. CALCULATED INFRARED SPECTRA

In the work reported here, all calculations of the geometric, electronic and vibrational properties of these compounds were carried out using GAUSSIAN-03/09 (Frisch et al. 2004) involving the highest precision density functional theory (DFT) formalism consistent with computational efficiency. In this case, this utilized a Lee-Yang-Parr correlation (Lee, Yang & Parr 1988) together with the three parameter exchange functional developed by Becke 1988 at the B3-LYP/6-311G* level of theory. For comparison with experiment, and to compensate in part for anharmonic effects, energies of infrared transitions as calculated from DFT were scaled by a factor 0.967. This factor was determined from a comparison between our calculated and laboratory energies measured for coronene, $C_{24}H_{12}$. The latter were obtained from Ne matrix data (Joblin et al 1995). Our scaling factor is consistent with the analysis of Fairchild et al. 2009.

We consider two groups of molecules classified as follows: molecules based on planar $C_{24}$ and small carbon cage molecules formed from pairs of dehydrogenated PAH. As noted above, cage molecules without pentagonal groups are of comparable stability to fullerenes for many small carbon clusters, and might then be expected to occur along with $C_{60}$ and $C_{70}$ in astrophysical sources. A good example is the $C_{24}$ molecule which, since it is more stable than the fullerene having the same number of carbon atoms, would then be a good indicator of the possible presence of smaller carbon clusters. Simulated infrared spectra for representative molecules in each of these classes are shown in Figs. 1 & 2 while predicted wavelengths for the strongest features in these spectra are listed in Table 1. The structures of some of these molecules are shown in Fig. 3.

## 3. DISCUSSION

The high symmetry of some of the molecules considered here, most notably $C_{24}$, $C_{24}^+$, $C_{24}^-$, $(C_5)_2$ and $(C_6)_2$, is reflected in the simplicity of their infrared spectra (Figs. 1 & 2). $C_{24}$ is characterized by three primary features located near 6.59, (9.76, 9.85) and (19.9, 20.3) μm. These three lines are repeated with small changes in wavelength in the spectrum of $C_{24}^+$ and, to a lesser extent, in that of $C_{24}^-$. Infrared spectra of $C_{24}^+$ and $C_{24}^-$ show additional features arising from C=C stretching vibrations at (5.07, 5.05) and 5.41 μm, respectively. The same three transitions at ~ 6.6, ~ 9.8 and ~ 20 μm are also observed, but less distinctly, in infrared spectra of substituted $C_{24}$ (Fig. 2) suggesting that these features are diagnostic of the presence of planar $C_{24}$. Significantly, this pattern is not observed in the infrared spectrum of $C_{24}H_{12}$ (Joblin et al. 1995).

Dimers formed from stacked pairs of dehydrogenated PAH molecules (Fig. 3) are characterized by strong (~ 7-8 eV) inter-layer bonds (Kuzmin & Duley 2010). The inter-layer bond in the $C_{24}$ dimer is ~ 26.9 eV. However, the strain in these structures implies that larger dimers such as $(C_{24})_2$ are less stable than the fullerene conformer (Wu et al. 2004). Infrared spectra (Fig. 2) of these structures, with the exception of $(C_5)_2$ and $(C_6)_2$, are less distinctive than those of $C_{24}$ and its derivatives. $(C_5)_2$ gives rise to features at 9.66, (16.2, 16.4), 19.2, and 26.5 μm, while the infrared spectrum of $(C_6)_2$ has its strongest bands at 8.34, (10.1, 10.2), 15.7 and 20.7 μm. The $C_{24}$ dimer, if it exists, would produce features at ~ 6.6, ~ 7.4, 10.6, 15.2 and 19.5 μm.

Detection of any of these molecules in sources that contain $C_{60}$ or other fullerenes will be complicated because some transitions (Table 1 & 2) overlap with the spectral features of this molecule. A full list of all infrared active vibrational modes in $C_{60}$ has been reported by Menendez & Page 2000. Inspection of these data show that the band at ~ 6.6 μm band predicted for $C_{24}$ based molecules falls into this category as it overlaps with a transition of $C_{60}$ at this wavelength. Features at 8.34 and ~10.1 μm in the calculated spectrum of $(C_6)_2$ also overlap with $C_{60}$ bands with approximately similar wavelengths. Separation of $C_{60}$ bands from those of smaller carbon clusters will be possible only when more precise wavelengths have been obtained either from laboratory measurements or from theoretical simulations using DFT calculations carried out at a higher level of theory.

## 4. CONCLUSIONS

Subject to the intrinsic limitations of DFT (Fairchild et al. 2009), we have simulated infrared spectra of a number of small carbon clusters, including $C_{24}$ and several of its derivatives for comparison with astronomical spectra. Planar $C_{24}$, which is more stable than its fullerene conformer, is characterized by a set of three spectral lines at ~ 6.6, ~ 9.8 and ~ 20.1 μm. Other small dehydrogenated carbon clusters including $(C_5)_2$ and $(C_6)_2$ have relatively simple spectra that may facilitate their detection in sources that show spectral features associated with fullerenes.

This work was supported by a grant from the NSERC and was made possible by the facilities of the Shared Hierarchical Academic Research Computing Network (SHARCNET: www.sharcnet.ca).

Table 1. Calculated wavelengths and energies of the strongest infrared transitions in spectra of molecules based on planar $C_{24}$.

| $C_{24}$ | | $C_{24}^+$ | | $C_{24}^-$ | | $C_{24}CH$ | | $C_{24}C_7H$ | |
|---|---|---|---|---|---|---|---|---|---|
| Energy ($cm^{-1}$) | λ (μm) | Energy ($cm^{-1}$) | λ (μm) | Energy ($cm^{-1}$) | λ (μm) | Energy ($cm^{-1}$) | λ (μm) | Energy ($cm^{-1}$) | λ (μm) |
| 2040 | 4.90 | 1982 | 5.05 | 1847 | 5.41 | 3104 | 3.22 | 3356 | 2.990 |
| 1518 | 6.59 | 1973 | 5.07 | 1393 | 7.18 | 3095 | 3.23 | 3346 | 2.989 |
| 1025 | 9.76 | 1518 | 6.59 | 1015 | 9.85 | 1537 | 6.51 | 3336 | 2.997 |
| 1015 | 9.85 | 1509 | 6.63 | 1006 | 9.94 | 1528 | 6.55 | 2021 | 4.95 |
| 503 | 19.9 | 1499 | 6.67 | 784 | 12.75 | 1518 | 6.59 | 1528 | 6.54 |
| 493 | 20.3 | 1257 | 7.96 | 774 | 12.9 | 1025 | 9.76 | 1508 | 6.63 |
| | | 1247 | 8.02 | 580 | 17.2 | 716 | 14.0 | 1480 | 6.76 |
| | | 1005 | 9.95 | 493 | 20.3 | 638 | 15.7 | 1035 | 9.66 |
| | | 996 | 10.0 | 455 | 22.0 | 493 | 20.3 | 1015 | 9.85 |
| | | 754 | 13.3 | 377 | 26.5 | 416 | 24.0 | 599 | 16.7 |
| | | 512 | 19.5 | 309 | 32.4 | 387 | 25.8 | 552 | 18.1 |
| | | 503 | 19.9 | 300 | 33.3 | 377 | 26.5 | 513 | 19.5 |
| | | | | | | | | 503 | 19.9 |

Table 2. Calculated wavelengths and energies of the strongest infrared transitions in spectra of carbon cage molecules based on pairs of dehydrogenated PAH molecules.

| $(C_5)_2$ | | $(C_6)_2$ | | $(C_{10})_2$ | | $(C_{14})_2$ | | $(C_{24})_2$ | |
|---|---|---|---|---|---|---|---|---|---|
| Energy $(cm^{-1})$ | λ (μm) | Energy $(cm^{-1})$ | λ (μm) | Energy $(cm^{-1})$ | λ (μm) | Energy $(cm^{-1})$ | λ (μm) | Energy $(cm^{-1})$ | λ (μm) |
| 1035 | 9.66 | 1199 | 8.34 | 1238 | 8.08 | 1363 | 7.34 | 1576 | 6.35 |
| 909 | 11.0 | 1151 | 8.69 | 1228 | 8.14 | 1267 | 7.89 | 1518 | 6.59 |
| 618 | 16.2 | 986 | 10.1 | 1141 | 8.76 | 1218 | 8.21 | 1508 | 6.63 |
| 609 | 16.4 | 977 | 10.2 | 1044 | 9.58 | 1080 | 9.26 | 1364 | 7.33 |
| 522 | 19.2 | 870 | 11.5 | 899 | 11.1 | 1151 | 8.69 | 1354 | 7.39 |
| 377 | 26.5 | 793 | 12.6 | 658 | 15.2 | 1102 | 9.07 | 1344 | 7.44 |
| | | 638 | 15.7 | 648 | 15.4 | 1044 | 9.58 | 986 | 10.1 |
| | | 522 | 19.2 | 560 | 17.9 | 957 | 10.45 | 948 | 10.55 |
| | | 484 | 20.7 | 561 | 17.8 | 948 | 10.55 | 841 | 11.9 |
| | | | | | | 880 | 11.4 | 687 | 14.55 |
| | | | | | | 870 | 11.5 | 658 | 15.2 |
| | | | | | | 803 | 12.45 | 513 | 19.5 |
| | | | | | | 793 | 12.6 | 348 | 28.7 |
| | | | | | | 735 | 13.6 | | |
| | | | | | | 667 | 15.0 | | |
| | | | | | | 619 | 16.15 | | |
| | | | | | | 541 | 18.5 | | |
| | | | | | | 455 | 22.0 | | |
| | | | | | | 416 | 24.0 | | |
| | | | | | | 406 | 24.6 | | |

FIGURE CAPTIONS

Fig. 1. Calculated infrared spectra of molecules based on planar $C_{24}$. Planar $C_{24}$ has greater stability than fullerenes with the same number of carbon atoms (Kent et al. 1998). Symmetries and multiplicities of the ground state are: $C_{24}$ ($D_{6h}$, singlet), $C_{24}^+$ ($C_{2h}$, doublet), $C_{24}^-$ ($C_{2v}$, doublet), $C_{24}CH$ ($C_s$, doublet), $C_{24}C_7H$ ($C_s$, doublet).

Fig. 2. Calculated infrared spectra of carbon cage molecules based on pairs of dehydrogenated PAH molecules. Due to their intrinsic stability, these molecules may also be present in regions where fullerenes are detected. Symmetries and multiplicities of the ground state are: $(C_5)_2$ ($C_2$, singlet), $(C_6)_2$ ($C_{2h}$, singlet), $(C_{10})_2$ ($C_{2h}$, singlet), $(C_{14})_2$ ($C_2$, singlet), $(C_{24})_2$ ($S_6$, singlet).

Fig. 3. Structures of some of the molecules whose spectra appear in Figs. 1 & 2.

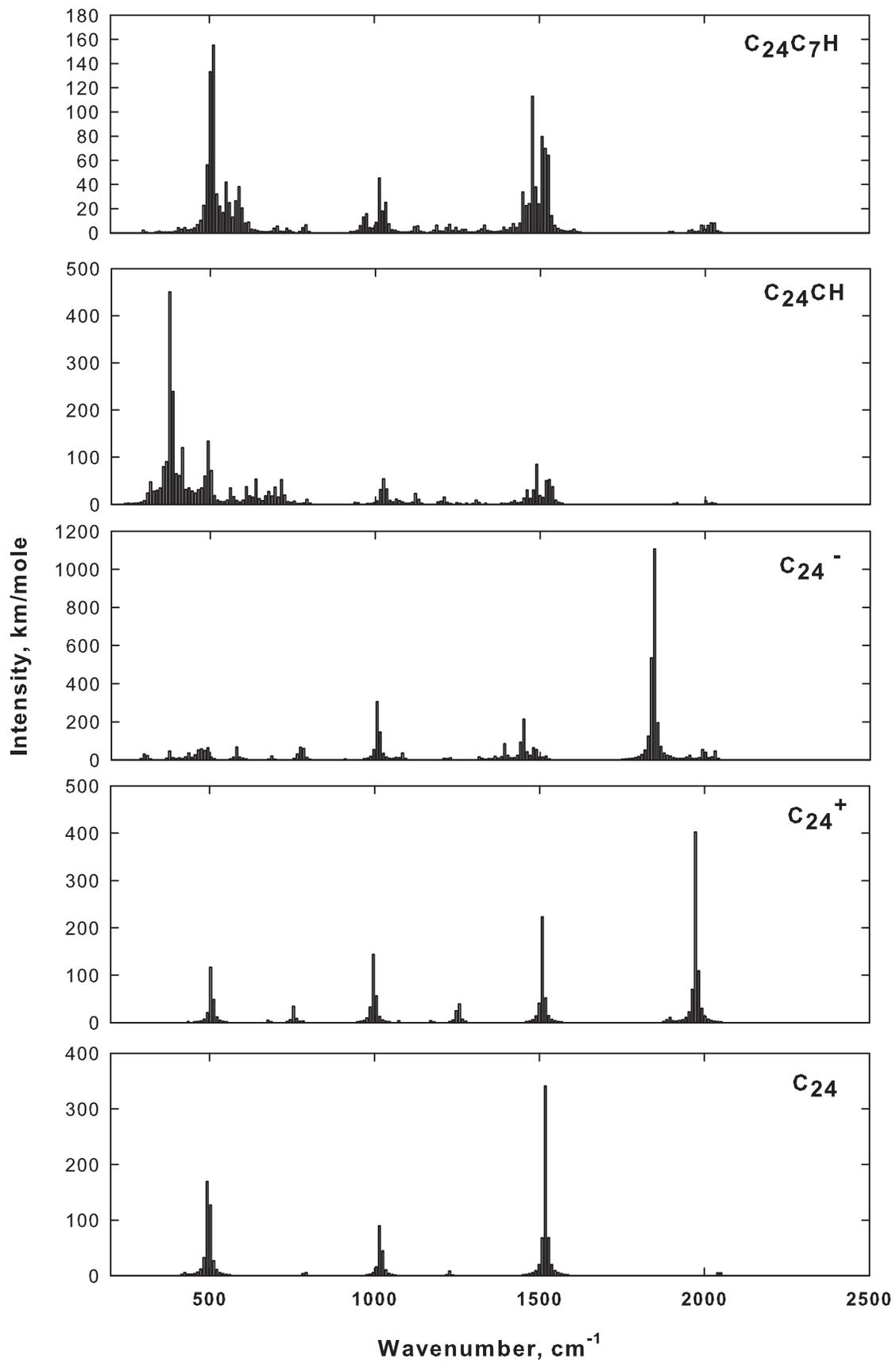

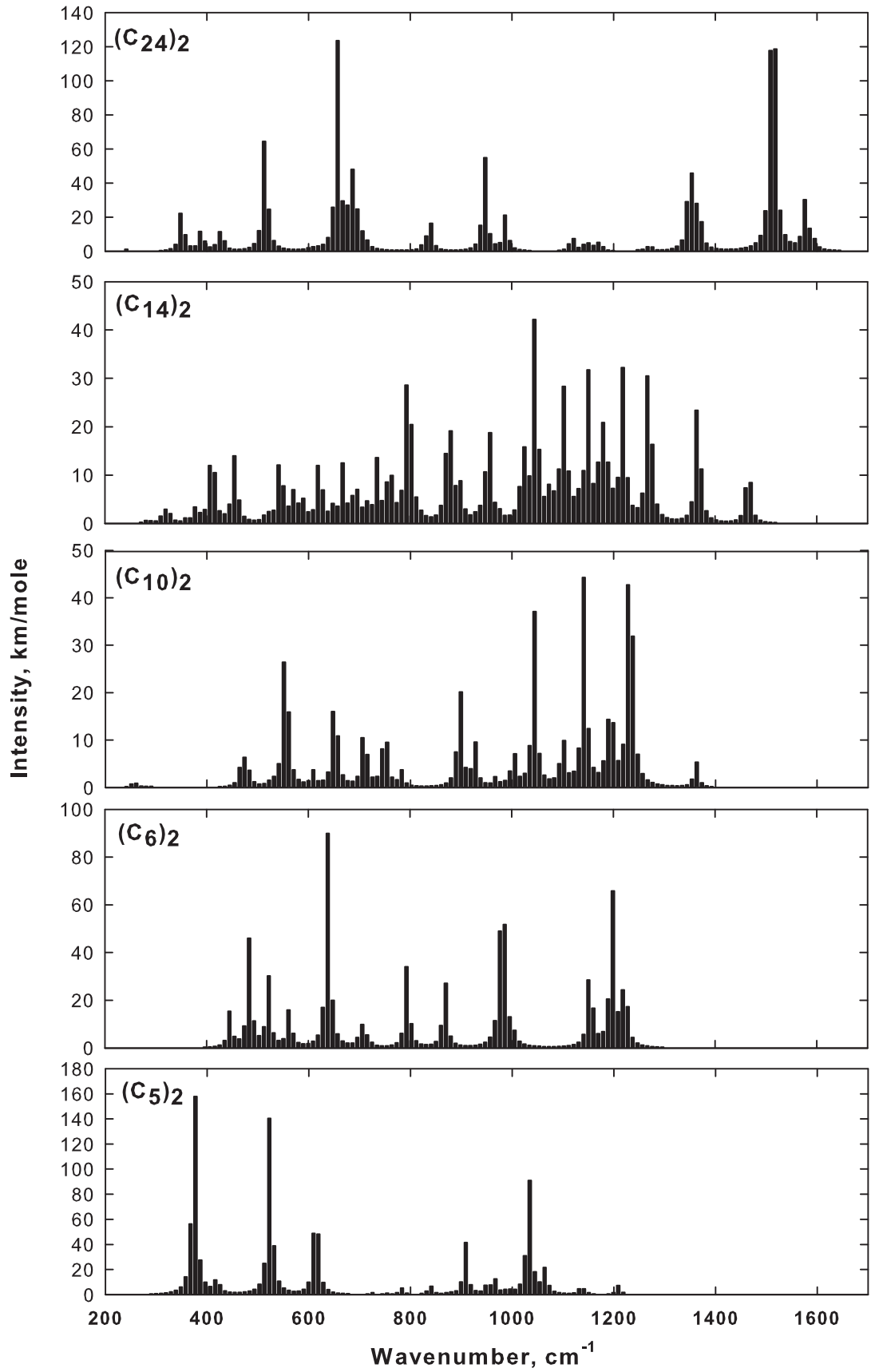

**(C$_5$)$_2$** 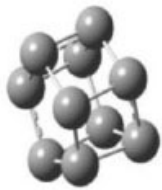

**(C$_6$)$_2$** 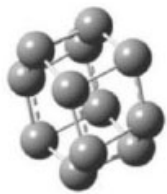

**(C$_{10}$)$_2$** 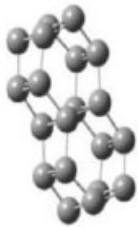

**(C$_{14}$)$_2$** 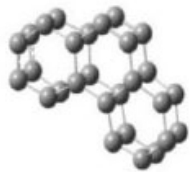

**(C$_{24}$)$_2$** 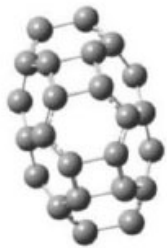

**C$_{24}$CH** 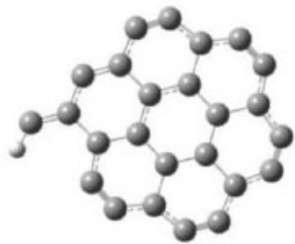